\documentclass[reprint,aps,pra,superscriptaddress,twocolumn]{revtex4-1}

\usepackage{amsmath}
\usepackage{graphicx}
\usepackage{hyperref}

\begin{document}


\title{Characterizing 
 $d-$dimensional 
quantum channels by means of quantum process tomography}

\author{J.~J.~M.~Varga}
\email[]{miguel@df.uba.ar}
\affiliation{Universidad de Buenos Aires, Facultad de Ciencias Exactas y Naturales, Departamento de F\'isica, Buenos Aires, Argentina}

\author{L.~Reb\'on}
\affiliation{Departamento de F\'isica, IFLP, Universidad Nacional de La Plata, C.C. 67, 1900 La Plata, Argentina}

\author{Q.~Pears~Stefano}
\affiliation{Universidad de Buenos Aires, Facultad de Ciencias Exactas y Naturales, Departamento de F\'isica, Buenos Aires, Argentina}

\author{C.~Iemmi}
\affiliation{Universidad de Buenos Aires, Facultad de Ciencias Exactas y Naturales, Departamento de F\'isica, Buenos Aires, Argentina}




\date{\today}

\begin{abstract}

In this work we propose a simple optical architecture, based on phase-only programmable spatial light modulators, in order to characterize general processes on photonic spatial quantum systems in a $d>2$ Hilbert space. We demonstrate the full reconstruction of typical noises affecting quantum computing, as amplitude shifts, phase shifts, and depolarizing channel in dimension $d=5$. We have also reconstructed simulated atmospheric turbulences affecting a free-space transmission of qudits in dimension $d=4$. In each case, quantum process tomography (QPT) was performed in order to obtain the matrix $\chi$ that fully describe the corresponding quantum channel, $\mathcal{E}$. Fidelities between the states experimentally obtained after go through the channel and the expected ones are above $97\%$.
\end{abstract}


\maketitle

In order to transmit quantum information along a communication channel or to study, for example, the dynamic of a quantum system, it is necessary to implement quantum operations on the degrees of freedom used to these purposes. A crucial point to make progress in this direction is to have a reliable method to characterize quantum devices, a task that it is possible to carry out by means of quantum process tomography (QPT) techniques \cite{PhysRevA.77.032322}.
For example, the analysis of the performance of a quantum communication channel enables to find the best alternative to protect the information against noise or to develop precise quantum error correction protocols \cite{Simon13}, allowing to improve the efficiency of the quantum communication.
More generally, QPT is a method for experimentally determining the unknown dynamics (open or close) of a quantum system under a large class of quantum operations including quantum algorithms, quantum channels, noise processes, and measurements \cite{ParisBook}.
  
Due to its favorable characteristics photonic systems raise as a suitable platform for quantum communications. Controllable operations on photonic quantum states has
been successfully demonstrated using
the polarization degree of freedom to codified the state \cite{kok2007linear}.
While these states are relatively simple to manipulate they only allow the realization of two-level
systems. Otherwise, higher dimensional quantum states,
namely qudits, can be used to increase the
quantum complexity without increasing the number of
particles involved. For this purpose, the discretized transverse momentum-position of single photons \cite{Neves2005} has become one of the main alternatives to codify $d$-level quantum systems. These photonic spatial qudits, usually called \textit{slit states}, are defined when photons are made to pass through a complex aperture with $d$ slits which set the qudit dimension. They have proven to be useful for several applications in quantum information science \cite{etcheverry2013,canas2014,matoso2016}. In this context, programmable optical devices, as spatial light modulators (SLMs), are used for state engineering and characterization \cite{lima2011,varga2017}. Only recently these devices have been introduced to implement more general quantum operations in slit qudits. In Ref. \cite{Padua2015} Marques et. al. used a SLM as a dissipative optical device to implement amplitude and dephasing  damping dynamics in $d=3$ and $d=4$. State transformations of qudits, encoded in the Gaussian spatial modes of the photon state, was proposed in Ref. \cite{Padua2017} and implemented for qutrits in Ref. \cite{Padua18}. However, still missing an experimental implementation of QPT for such systems. In fact, QPT has been applied mainly to qubit systems	 \cite{Howard06,Mataloni10}.

In this letter, we present for the first time the realization of QPT in slit qudits. To this end we propose an optical architecture in which a phase- only SLM is used to mimic the transformation of an initial quantum state $\rho_{in}$ through a quantum channel $\mathcal{E}$, so that, $\rho_{in} \overset{\mathcal{E}}{\rightarrow }\rho_{out}=\mathcal{E}(\rho_{in})$. A second SLM is used to implement the set of projective measurements for a complete characterization of the final quantum state. 
As shown hereafter, with this architecture we are able to apply the standard quantum process tomography (SQPT) technique \cite{NielsenChuang97} for characterizing processes in any dimension $d$.

The general procedure for SQPT can be summarized as follow: A quantum process can be described by a completely positive linear map $\mathcal{E}$. In the so-called, \textit{operator-sum representation} or \textit{Kraus decomposition}, it gives the dynamics of a quantum system by means of 
$\mathcal{E}(\rho)=\sum_{k}E_k\rho E_k^{\dag}$,
where $E_k$ are operators from the space of $d\times d$ density matrices in itself, and satisfy the relation $\sum_{k}E_kE_k^{\dag}\leq\hat{1}$. It can be written equivalently as 
$\mathcal{E}(\rho)=\sum_{m,n}~A_m~\rho~A_n^{\dag}~\chi_{mn}$,
where $\{A_i\}_{i=0}^{d^2-1}$ is a fix basis of operators. Thus, $\mathcal{E}$ can be completely described by a $d^2 \times d^2$ complex matrix $\chi$, once the operators $A_i$'s are chosen. Experimentally, the states $\{\rho_0, \rho_1,...,\rho_{d^2-1}\}$, forming a basis for the space of density matrices, are prepared and the unknown process $\mathcal{E}$ is applied. The output state $\mathcal{E}(\rho_i)$ is determined, for each input $\rho_i$, by quantum state tomography (QST). Since $\mathcal{E}$ is linear, the measurement results $\{\mathcal{E}(\rho_0), \mathcal{E}(\rho_1),...,\mathcal{E}(\rho_{d^2-1})\}$ are enough to find the action of the process on any state $\rho$, i.e., after linear algebraic calculations, the matrix $\chi$ is obtained from the set of experimental data.

\begin{figure}[htbp]
\centering
\includegraphics[width=0.9\linewidth]{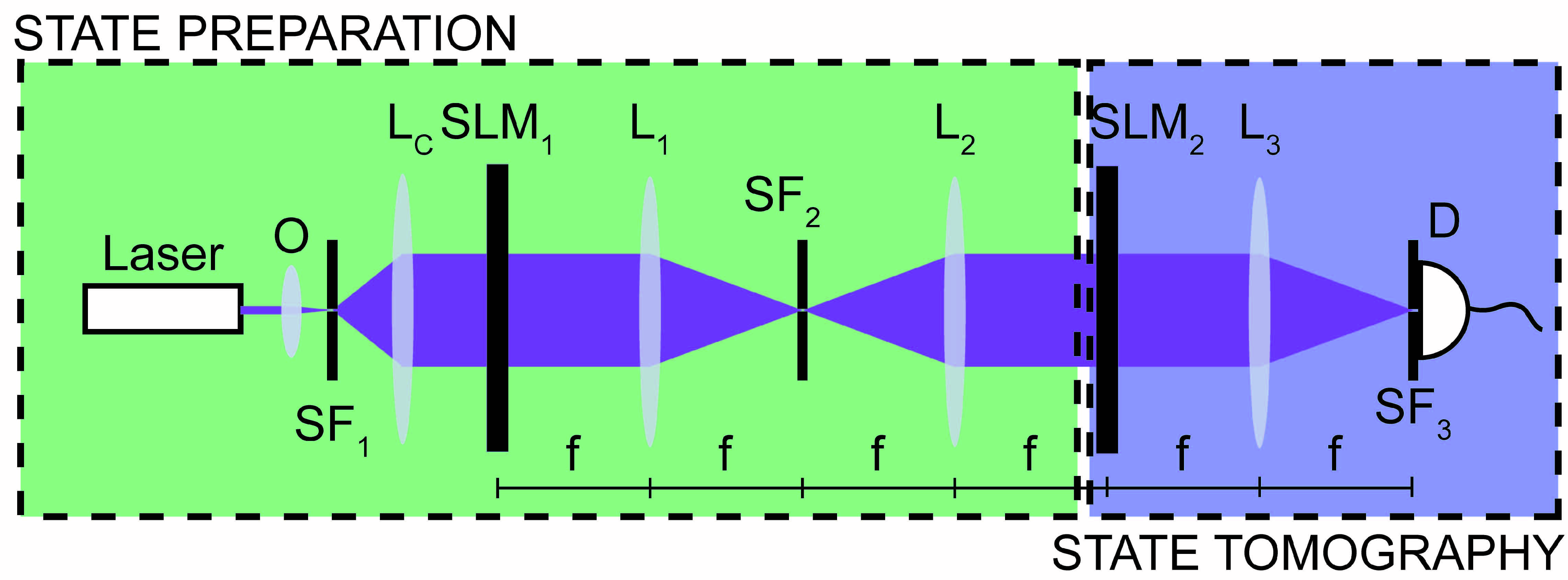}
\caption{Experimental setup. $O$ is an expansor, $\text{SF}_i$ are spatial
filters, $\text{L}_i$ are lenses with a focal distance $f$, $\text{SLM}_i$ are spatial light modulators, and $D$ is a single pixel detector.} \label{fig:000_setup}
\end{figure}

The experimental implementation is based on the setup schematically shown in Fig.~\ref{fig:000_setup}. The slit states are generated and reconstructed after QST following the methods described in Ref.~\citep{varga2017}. 
The first part of the setup, used for state preparation, consists of a cw 405 nm single mode laser diode whose transverse spatial profile is proportional to the transverse probability amplitude of a single-photon field. The attenuated laser beam was spatially filtered and collimated. Thus, the beam transverse profile impinges on $\text{SLM}_1$ with a planar wave with approximately constant phase and amplitude distribution in the region of interest. The required pure phase modulation was provided by a Sony liquid crystal television panels model LCX012BL  in combination of polarizers and wave plates that provide the adequate state of light polarization to reach a phase modulation near to $2\pi$@405 nm \cite{Marquez2001}. With this architecture, we can generate pure spatial qudits, $|\psi\rangle=\frac{1}{\sqrt{d}}\sum_{\ell=0}^{d-1} c_{\ell}|\ell\rangle$, with arbitrary complex coefficients $c_{\ell}=\beta_{\ell}~ e^{~ i\phi_{\ell}}$. The coefficient modulus $\beta_{\ell}$, is given by the phase modulation of the diffraction gratings displayed on each of the $d$-slit regions, while the argument $\phi_{\ell}$ is defined by adding a constant phase value. The spatial filter $\text{SF}_2$ is used to select the first order diffracted by the mentioned gratings in such a way that on the back focal plane of lens $\text{L}_2$ it is obtained the complex distribution that represents the quantum state of the desired spatial qudit, $|\psi\rangle$. Even more, since the complex modulation capabilities of our architecture allows us to \textit{dynamically modify} both, the phase and the amplitude of each slit, we are able to simulate the evolution of an initial pure state to a final arbitrary mixed state \citep{varga2017}. Then, this first SLM is used to simulate the action of a quantum process $\mathcal{E}$ on each of the basis states $\{\rho_i\} _{i=0}^{d^2-1}$ and, after filtering, what we obtain is just $\mathcal{E}(\rho_i)$. 

A second modulator
$\text{SLM}_2$ is placed in the
front focal plane of $\text{L}_2$. By following the
same method described previously, we represented on it the reconstruction basis used to implement the QST process. The first SLM ($\text{SLM}_1$) is imaged onto the second one ($\text{SLM}_2$),
while a spatial filter $\text{SF}_3$ and a single pixel detector placed at the back focal plane of $\text{L}_3$ are used to select and measure the intensity in the center of the interference pattern produced by the slits. This intensity is proportional to the probability of projecting the state defined by $\text{SLM}_1$
onto the state defined by $\text{SLM}_2$ \cite{lima2011}. 
The projections of each of the unknown states after process, $\mathcal{E}(\rho_i)$, are performed onto the informational complete set of mutually unbiased bases (MUBs).
Then, this set of measurements results is all that we need to determined the matrix $\chi_\mathcal{E}$ that characterize the process $\mathcal{E}$.

In order to test the performance of the setup for QPT, we have simulated and reconstructed five different quantum process, which are particularly relevant in applications such as quantum computing 
or quantum communications. 
We start with a first group of processes (amplitude shifts (AS), phase shifts (PS), amplitude-phase shifts (APS) and depolarizing channel (DC)) which are suitable models for different kind of errors in a quantum computer ~\citep{Pirandola08}. Since their decomposition in terms of Kraus operators are known, 
they can be controllably implemented in the laboratory in an easy way. In addition, the experimental results are straightforward to interpret and compare with the theoretical results. As we mentioned previously, in the operator-sum representation the input-output relation can be written as $\mathcal{E}(\rho)=\sum_{k}E_k\rho E_k^{\dag}$. We have define $E_k\equiv\sqrt{p_{\nu\alpha}}\tilde{E}_{\nu}^{(\alpha)}$, where $\{p_{\nu\alpha}\}$ is the set of parameters that represent the weight of each Kraus operator $\tilde{E}_{\nu}^{(\alpha)}$ $(\sum_{\nu\alpha}p_{\nu\alpha} =1)$. The explicit form of these operators, for the processes to be considered here, is
\begin{eqnarray}
\tilde{E}(X)_{\nu}^{(\alpha)}  =
	\left \{
	\begin{array}{rcl}
	\hat{1}~~~~~~~~&,& ~ \nu,\alpha=0, \\G_{\nu}^{(\alpha)\dag}~X~G_{\nu}^{(\alpha)}&,&  ~ \nu,\alpha=0,1,...,d-1~,~\alpha>\nu
	\end{array}
	\right.
\end{eqnarray}
where $G_{\nu}^{(\alpha)}$ is a 2 x $d$ matrix whith elements $(G_{\nu}^{(\alpha)})_{i,j}=\delta_{i1}\delta_{j\nu}+\delta_{i2}\delta_{j\alpha}$, and $X$ is one of the 2 x 2 Pauli matrices, depending on the process (AS $\rightarrow\sigma_x$, PS $\rightarrow\sigma_z$, APS $\rightarrow\sigma_y$). For a DC with probability $1-p$ that the system remains in the state $\rho$, and a probability $p$ that a general error occurs, the Kraus decomposition is given by
\begin{eqnarray}
\mathcal{E}_{DC}(\rho)=(1-p)\rho +\frac{p}{3}\sum_{r=x,y,z}\sum_{\nu\alpha}\tilde{E}(\sigma_r)_{\nu}^{(\alpha)}~\rho ~\tilde{E}(\sigma_r)_{\nu}^{(\alpha)\dag}.
\label{depolarizing}
\end{eqnarray}

\begin{figure}[htbp]
\centering
\includegraphics[width=0.85\linewidth]{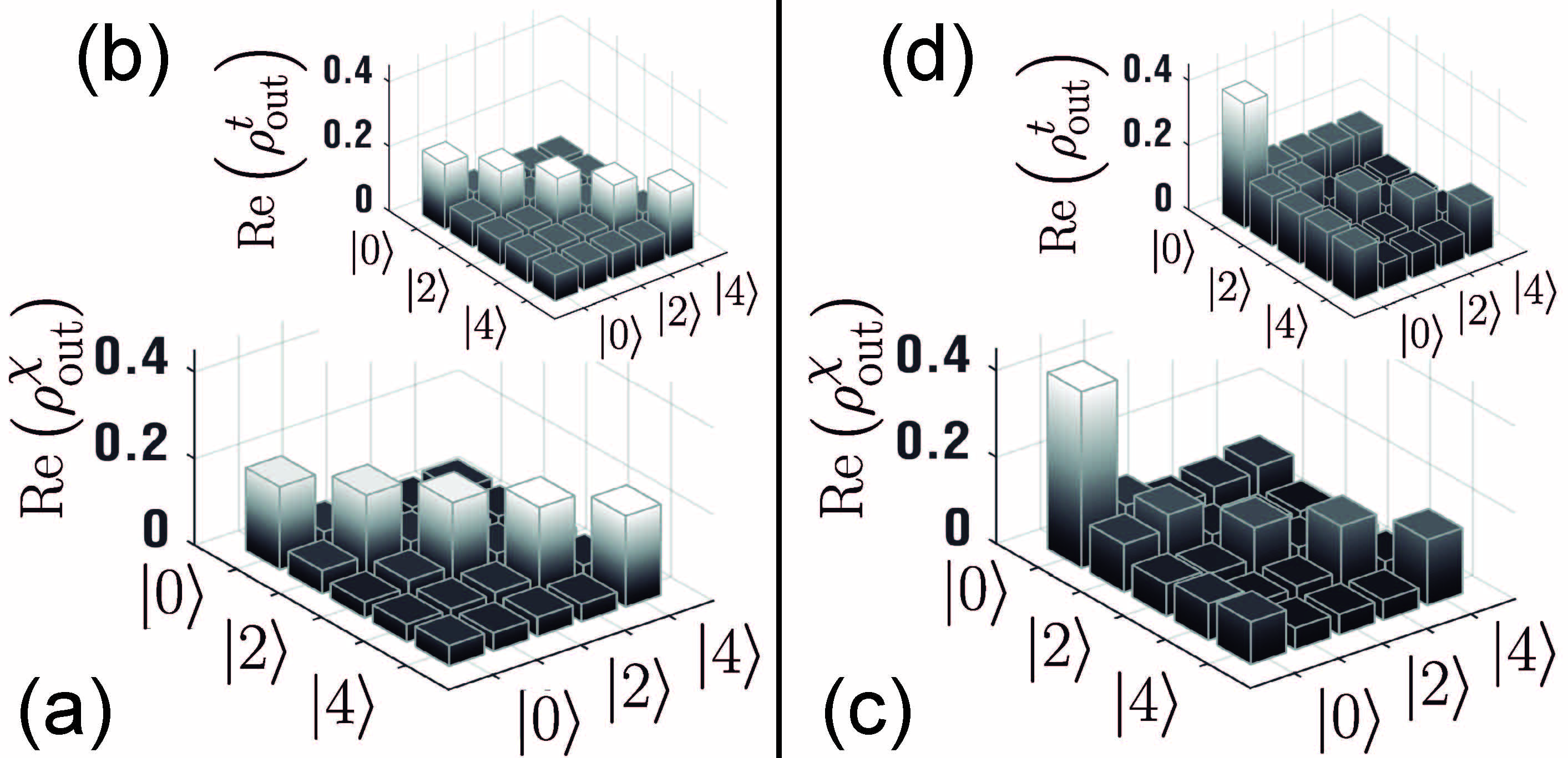}
\caption{{\it Amplitude-shift} (AS): Predicted density matrices $\rho_{out}=\mathcal{E}_{AS}(\rho_{in})$, for an uniform AS (left panel) and an uniform-respect-to-$|0\rangle$ AS (right panel), in $d=5$. Plots (b) and (d) show the density matrices of the outputs states, obtained by means of the Kraus decomposition of the process, while (a) and (c) are the corresponding density matrices obtained by means of the reconstructed matrix of such a process, $\chi$, after SQPT. The input state is $|\psi\rangle=\frac{1}{\sqrt{5}}\sum_{\ell=0}^{4} |\ell\rangle$. The imaginary parts of the coefficients, not shown here, are zero or almost zero. 
} 
\label{fig:01_result_bit_flip}
\end{figure}

Bellow, we show the results obtained from the characterization of these processes. In Fig.~\ref{fig:01_result_bit_flip}, we can see the comparison between the predicted density matrices $\rho_{out}^{\chi}$ and $\rho_{out}^t$, for a given input state $\rho_{in}=|\psi\rangle\langle\psi|$ after an AS process. The first one is obtained by means of the reconstructed matrix of the process $\chi$ (Figs.~\ref{fig:01_result_bit_flip} (a) and \ref{fig:01_result_bit_flip} (c)), while the second is directly obtained from the theoretical Kraus decomposition (Figs.~\ref{fig:01_result_bit_flip} (b) and \ref{fig:01_result_bit_flip} (d)). The left (right) panel corresponds to an AS in dimension $d=5$, with 
probabilities $p_{\nu\alpha}=\frac{1}{\sqrt{11}}$ $\forall \nu, \alpha=0,...,d-1 (\alpha >\nu)$, and $p_{\nu\alpha}=\frac{1}{\sqrt{5}}\delta_{\nu0}$, respectively. In both cases the initial state was chosen to be $|\psi\rangle=\frac{1}{\sqrt{d}}\sum_{\ell=0}^{d-1} |\ell\rangle$. As figure of merit, we use the fidelity between the density matrices of these two output states, defined as $F\equiv F(\rho_{out}^t,\rho_{out}^{\chi})=Tr\sqrt{\sqrt{\rho_{out}^t}\rho_{out}^{\chi}\sqrt{\rho_{out}^t}}$, is $F=0.980$ ($F=0.989$). 
Ideally, it is desirable to have $F=1$.  
Additionally, we have calculated $F\equiv$, considering as input state each one of the states of the basis, $\{\rho_i\}_{i=0}^{d^2-1}$. We have obtained a mean value of $F_{m}=0.9821$ ($F_{m}=0.978$).
Similar fidelities were obtained for PS and APS processes in $d=5$ (See Fig.~\ref{fig:visualization1} and Fig~\ref{fig:visualization2} in Supplementary Material).    

\begin{figure}[htbp]
	\centering	\includegraphics[width=0.7\linewidth]{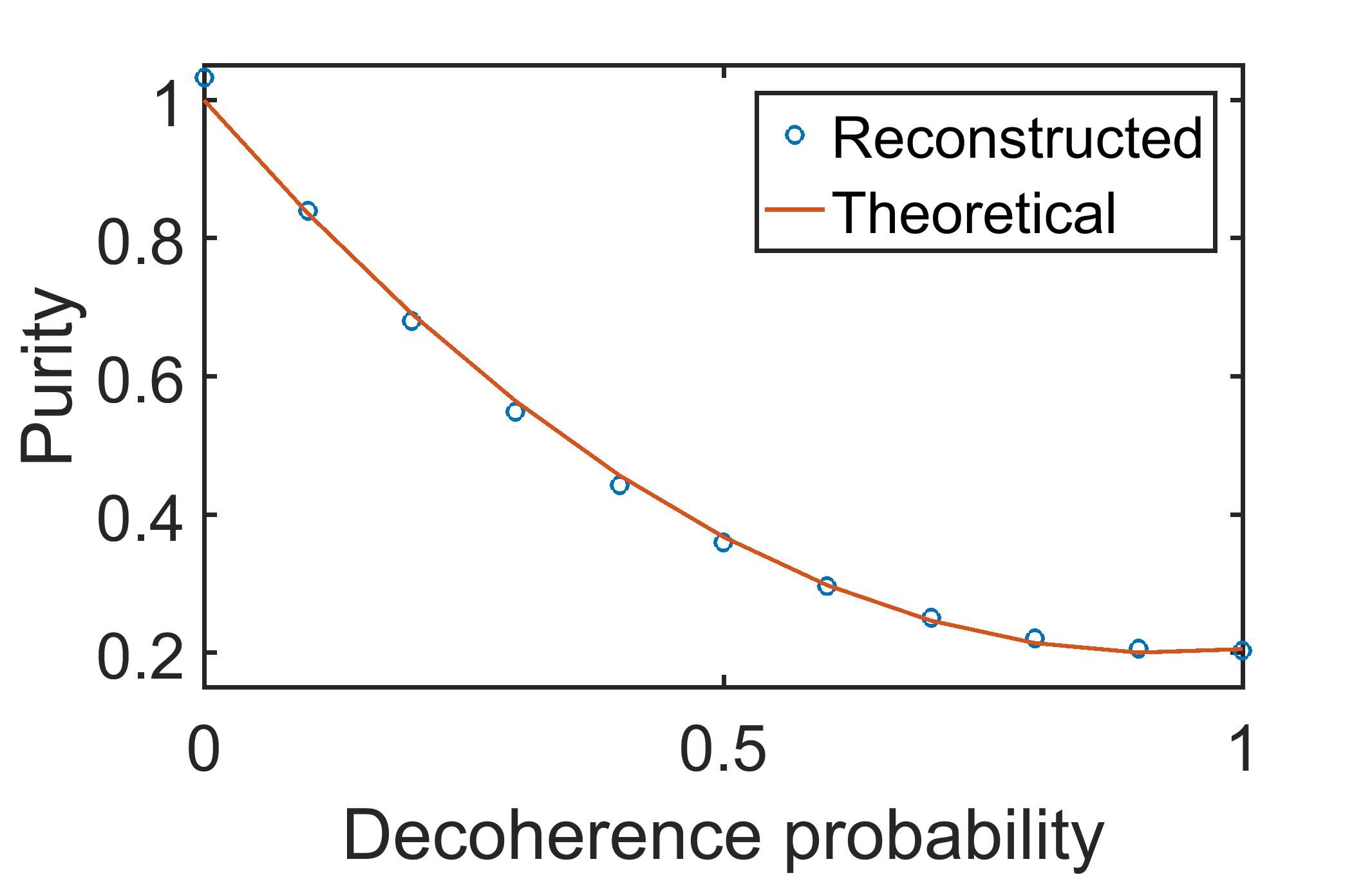}
\caption{{\it Depolarizing channel} (DC): Purity of the output state $\rho_{out}=\mathcal{E}_{DC}(\rho_{in})$, as a function of the decoherence probability $p$, given the input state $|\psi\rangle=\frac{1}{5}\sum_{\ell=0}^{4} |\ell\rangle$. Theoretical function (red line), and reconstructed after SQPT (blue circles).
} \label{fig:04_b_dep_ch_pur_vs_prob}
\end{figure}

We have simulated and characterized a quantum DC  
in dimension $d=5$, for different values of the depolarizing probability $p$ (see Eq.~\ref{depolarizing}). With $p$ as control parameter, it is possible to set the quantum coherence 
of the output state, $\rho_{out}$, for a given input state, $\rho_{in}$. In Fig.~\ref{fig:04_b_dep_ch_pur_vs_prob} we show the purity of $\rho_{out}$ $(P(\rho_{out})\equiv Tr\rho_{out}^2)$ when the input state is the pure state $|\psi\rangle=\frac{1}{\sqrt{d}}\sum_{\ell=0}^{d-1} |\ell\rangle$ $(P(\rho_{in})=1)$. The red line is the 
theoretical curve $P(\rho_{out}^t)~vs~p$, while the blue circles correspond to the values of this relation when $\rho_{out}=\rho_{out}^{\chi}$. Supporting the excellent agreement, the mean value of the fidelity, $F$, over the different evolutions is $F_{m}=0.990$.

Finally, we have performed the SQPT of a free - space communication channel over long distances, to study the effect of atmospheric turbulence (AT) on slit states. This topic is of great interest for free space quantum communications, where AT affects the quality of the transmitted information. By taking advantage of the modulation capabilities of programmable SLMs, we have created stochastic masks from the
superposition of normal random phase modes that follows the power laws dictated by the Kolmogorov statistic~\cite{10.2307/51980}. These power laws must be fulfilled in the amplitude of the modes and in the periodicity in time with which they modify their random phase. This ensures the self-similarity condition of turbulent fluids. In order to link the ATs simulated in the laboratory with real communicational situations, there are several empirical models that relate the intensity with the height $h$ above the sea level in which communication takes place~\cite{Valley:80}.

\begin{figure}[htbp]
\centering
\includegraphics[width=0.8\linewidth]{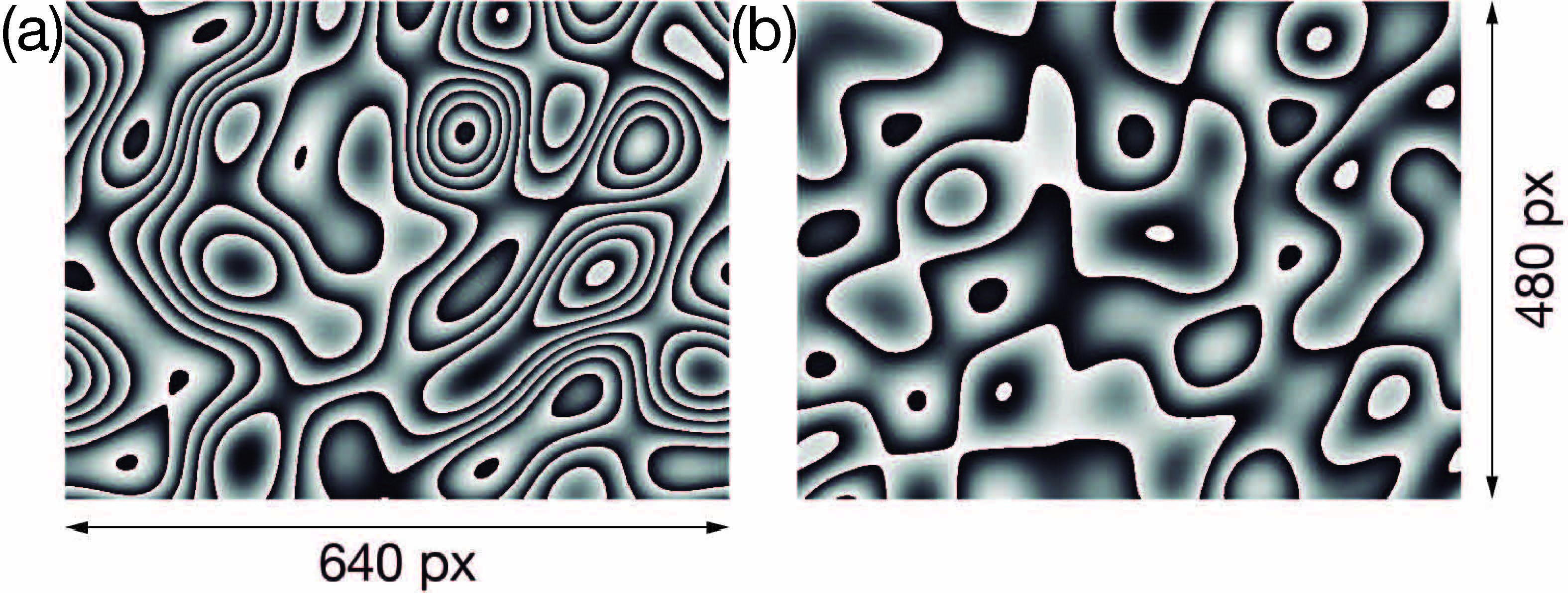}
\caption{Turbulence masks addressed to the SLM. There are a superposition of normal modes of random phases that follows a power law in time and amplitude. Case {\bf(a)} corresponds to a free path communication at $h=174$m above sea level ($L=500$m). Case {\bf(b)} corresponds to a free path communication at $h=647$m above sea level. } \label{fig:05_a_result_turb_mask}
\end{figure}

In Fig.~\ref{fig:05_a_result_turb_mask} we show two different turbulence masks implemented in our experiment. The gray levels represent the phase introduced by these masks, being white for 0 radians and black for $2\pi$ radians. The masks simulate a free path communication of distance $L=500$m  for the atmospherical conditions at $h=174$m (Figures~\ref{fig:05_a_result_turb_mask} (a)) and $h=647$m (\ref{fig:05_a_result_turb_mask} (b)) above sea level, respectively.

\begin{figure}[htbp]
\centering
\includegraphics[width=0.85\linewidth]{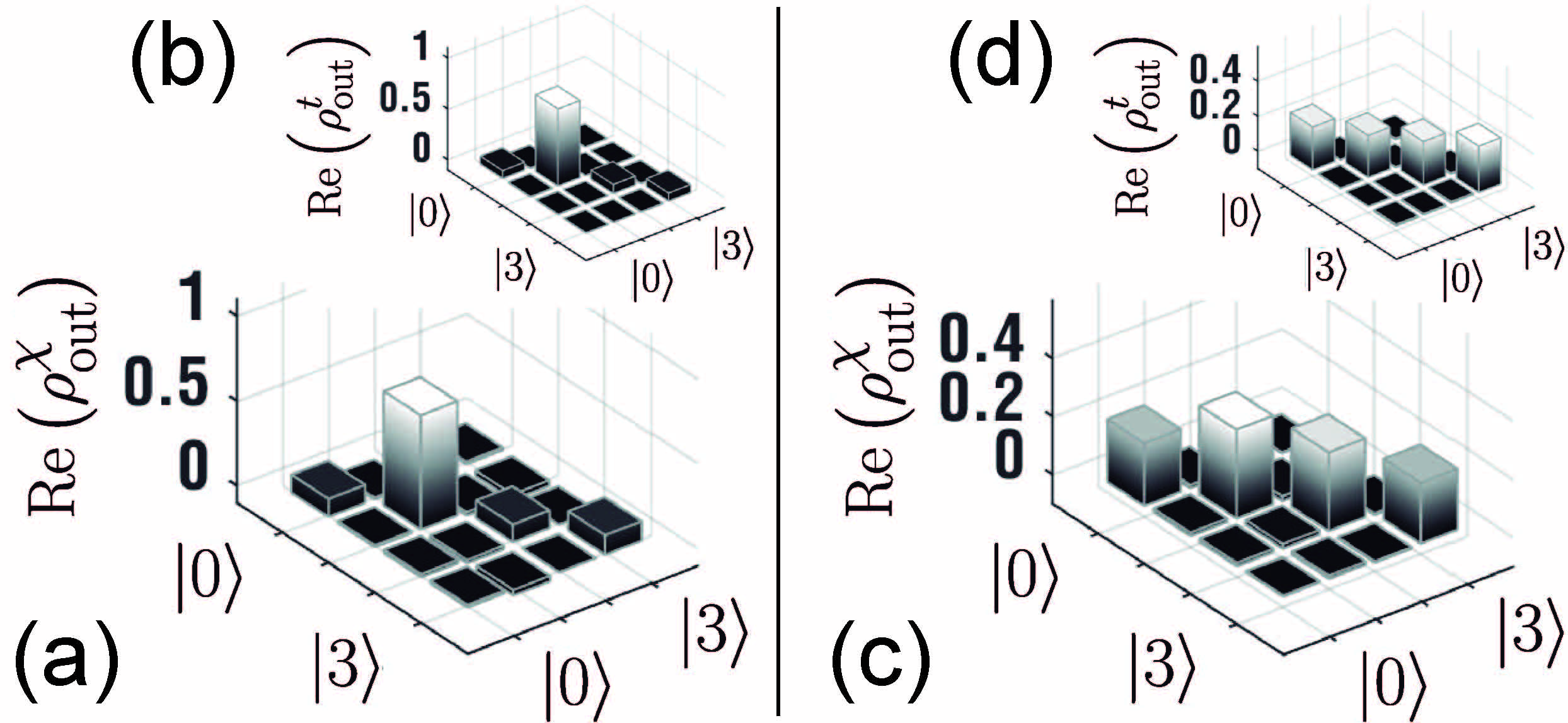}
\caption{{\it Atmospheric turbulence} (AT): Predicted density matrices  theoretical, $\rho_{out}^t$, and reconstructed $\rho_{out}^{\chi}$ after SQPT, for a turbulent free-communication channel, in $d=4$. The turbulence mask corresponds to case (a) in Fig.~\ref{fig:05_a_result_turb_mask}, for an input state $|\psi\rangle=|1\rangle$ (left panel), and $|\psi\rangle=\frac{1}{4}\sum_{\ell=0}^{3} |\ell\rangle$ (right panel). 
}
\label{fig:turb_matrix_pi-01}
\end{figure}

\begin{figure}[htbp]
\centering
\includegraphics[width=0.85\linewidth]{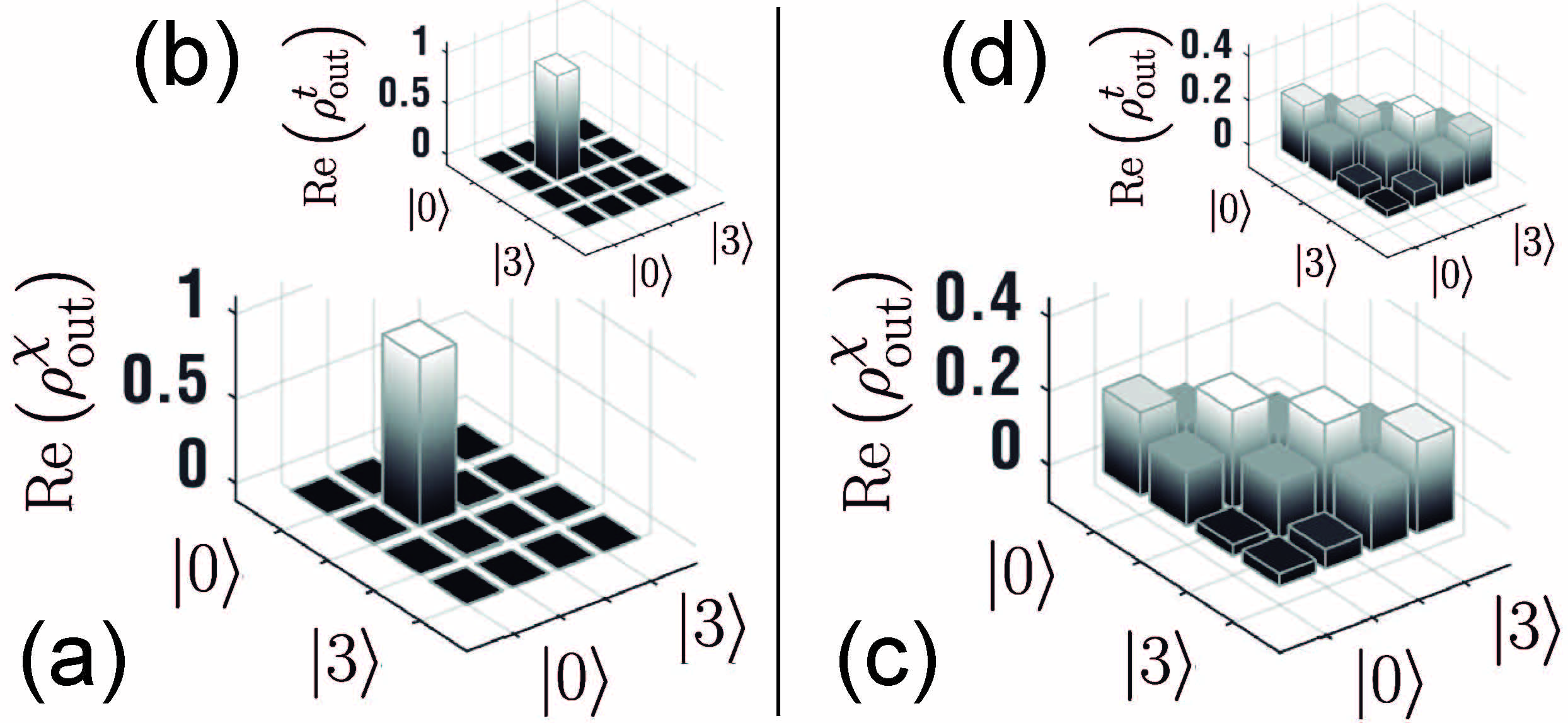}
\caption{Idem Fig.~\ref{fig:turb_matrix_pi-01} for a turbulence mask corresponding to case (b) in Fig.~\ref{fig:05_a_result_turb_mask}.
}
\label{fig:turb_matrix_pi_4-01}
\end{figure}

The characterization of these channels was carried out at the same time that we randomly varied, in time, the turbulence masks addressed on the first SLM. We have followed a similar procedure in Ref.~\cite{varga2017} to see the evolution of a quantum system, from a pure state to a final mixed state. The SQPT gives a convergence matrix $\chi$ after a random superposition of 500 masks. In Fig.~\ref{fig:turb_matrix_pi-01} we show the comparison between the predicted density matrices $\rho_{out}^{\chi}$ and $\rho_{out}^t$, after SQPT, for a turbulent free-communication channel, in $d=4$. The input state is $|\psi\rangle=|1\rangle$ in the left panel, and $|\psi\rangle=\frac{1}{4}\sum_{\ell=0}^{3} |\ell\rangle$  in the right panel. In both cases, the
process corresponds to a AT with the parameters as that represented in Fig.~\ref{fig:05_a_result_turb_mask} (a). The same comparison is shown in Fig.~\ref{fig:turb_matrix_pi_4-01} for the atmospherical conditions as that represented in Fig.~\ref{fig:05_a_result_turb_mask} (b). 

In Fig.~\ref{fig:turb_matrix_pi-01}, left panel, it is noticeable
the appearance of non null populations of the elements $|0\rangle$, $|2\rangle$ and $|3\rangle$. This phenomenon, called crosstalk, is common in several implementations of quantum communications. In the case of spatial states, the reason for this crosstalk is the deviation of the photons due to the strong phase variation on the optical path. On the other hand, in the right panel, it is evident that the coherences of the output state practically vanish. In fact the purity of the output state is $P=0.26$. This destruction of the coherence is due to the randomness of the mean phase in each slit, that generates null interference in the far field. In such a case, there is no information about the pre-channel state.

The same analysis is performed in Fig.~\ref{fig:turb_matrix_pi_4-01}. Unlike the previous case,  
we do not find crosstalk between the populations (left panel). This is due to the fact that the lower intensity of the AT does not disturb the optical path in a way that photons impact on zones corresponding to neighboring slits. Besides, this less intense AT, does not completely destroy the coherences between slits.
The right panel of Fig.~\ref{fig:turb_matrix_pi_4-01} shows a gradual decay of coherence as a function of the relative labels between the slits. The reason for this particular behavior is that, for elements of the spatial codification basis 
$\{|\ell\rangle\}_{i=0}^{d-1}$, more distant from each other, the phase difference introduced by the AT is 
greater.
This partial information that survives the channel, allows us to devise a method for recovering, after post-processing of data, the original input state. 
Our proposal implies finding an inverse process matrix $\Xi$, which allows recovering the
pre-channel states from the output states. 

We have chosen as fix basis of operators, $\{A_i\}_{i=0}^{d^2-1}$, the projectors $P_{i+j}=\vert i\rangle\langle j\vert$ for $i,j=0, 1,...,~d-1$.
Then, once the matrix $\chi$ has been found by means of SQPT, it is straightforward to obtain the inverse process matrix $\Xi$:
\begin{eqnarray}
\Xi&=&\left(%
\begin{array}{cccc}
 \Xi_{0}  & \Xi_{1} & \cdots & \Xi_{d-1}\\
 \Xi_{d}  & \Xi_{d+1} & \cdots & \Xi_{2d-1}\\
 \vdots  & \vdots & \cdots & \vdots\\
 \Xi_{(d-1)d}  & \Xi_{(d-1)d+1} & \cdots & \Xi_{d^2-1}
\end{array}%
\right),
\end{eqnarray}
where $\Xi_k$ is a $d\times d$ matrix with elements $(\Xi_k)_{i,j}=\chi_{k,~id+j}$ and $i,j=0,1,...d-1$. Notice that $\Xi$ is not the inverse of $\chi$,  
but a matrix that reverses the effects of the process ($\rho_{in}=\Xi~\rho_{out}$)

\begin{figure}[htbp]
\centering
\includegraphics[width=0.75\linewidth]{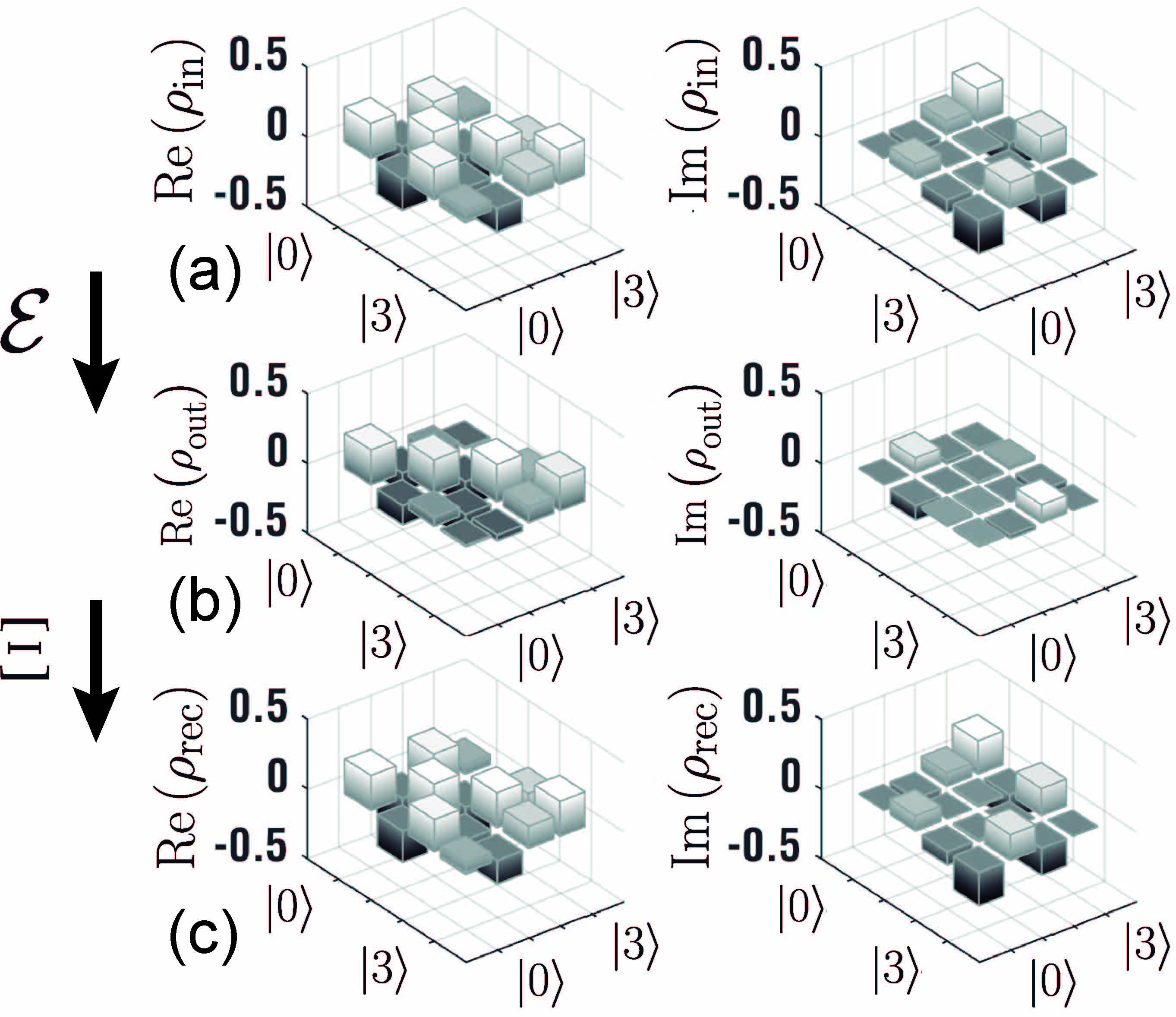}
\caption{\emph{State recovery:} {\bf(a)} Arbitrary initial state before going through a turbulent channel corresponding to case (b) in Fig.~\ref{fig:05_a_result_turb_mask}. {\bf(b)} Output state reconstructed by means of the process matrix $\chi$. {\bf(c)} Recovered state by means of the inverse process matrix, $\Xi$.
}
\label{fig:12_turb_recup-01}
\end{figure}

In Fig.~\ref{fig:12_turb_recup-01} it is displayed an example of this. We have chosen an input state $|\psi\rangle=\frac{1}{2}\sum_{\ell=0}^3e^{i\phi_\ell}|\ell\rangle$, with arbitrary $\phi_\ell$, shown in Fig.~\ref{fig:12_turb_recup-01} (a). After the AT-channel, the output state $\rho_\text{out}$ is shown in Fig.~\ref{fig:12_turb_recup-01} (b).
Finally, in Fig.~\ref{fig:12_turb_recup-01} (c), it can be observed that from the application of the matrix $\Xi$ to the output state, it is possible to recover the original state, intended to be communicated.

In conclusion, the proposed optical device has proved to be an useful and flexible tool to implement QPT in high-dimensional Hilbert spaces. We have carried out the reconstruction of noisy processes, typically related to quantum computing, and a simulation of AT, that usually affects the transmission of information in free space. For this last case we have proposed a method that, depending on the intensity of the turbulence, allows us to recover the initial information.
\vspace{20pt}

\begin{acknowledgments}
We thank P. Mininni and P. Cobelli for helpful
discussions. This work was supported by UBACyT 20020130100727BA and ANPCYT PICT 2014/2432.
\end{acknowledgments}
\vspace{-11pt}

\section*{Supplementary Material}
\begin{figure}[htbp]
\centering
\includegraphics[width=0.75\linewidth]{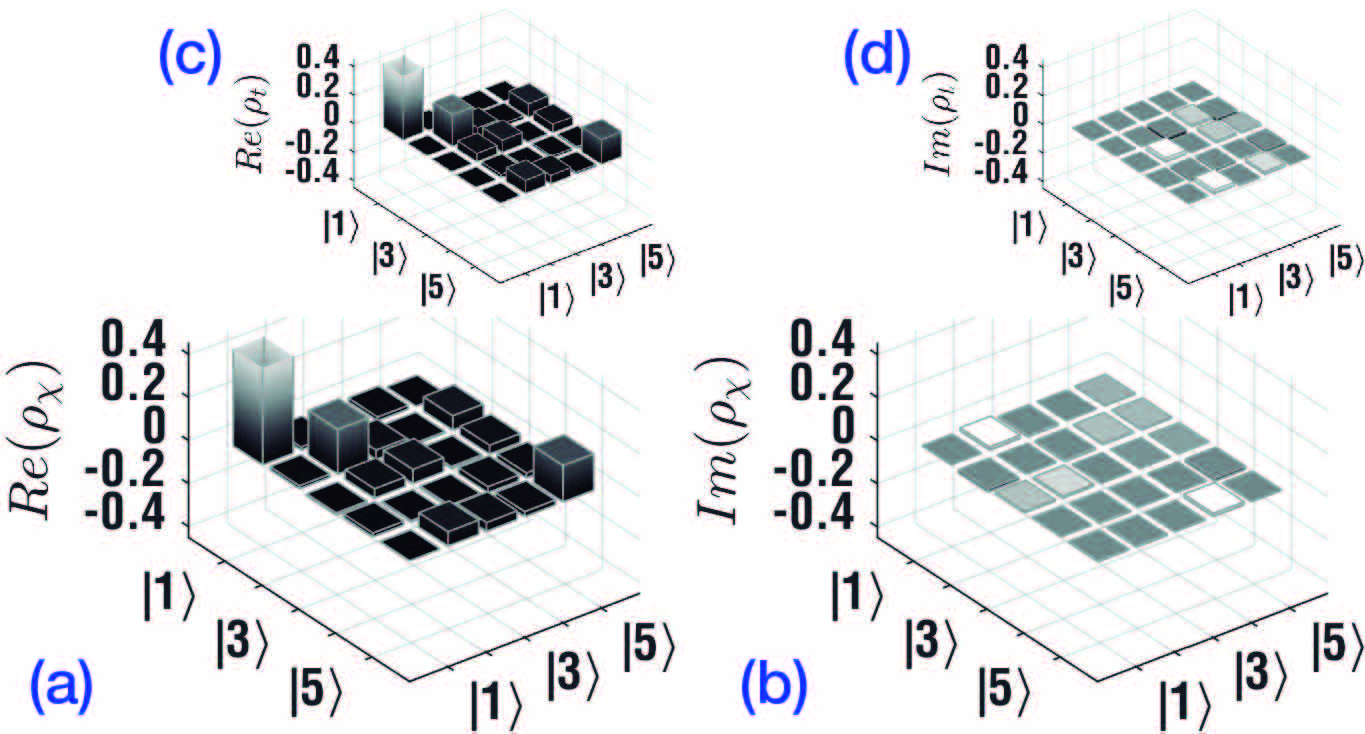}
\caption{{\it Phase-shift} (PS): Predicted density matrices $\rho_{out}=\mathcal{E}_{PS}(\rho_{in})$, for an uniform-respect-to-$|0\rangle$ PS, in $d=5$. Plots (c) and (d) show the real and imaginary parts of the density matrices of the outputs states, obtained by means of the Kraus decomposition of the process, $\mathcal{E}_{AS}$. Plots (a) and (b) are the real and imaginary parts corresponding to the density matrices obtained by means of the reconstructed matrix of such a process, $\chi$, after SQPT. The input state is a random state. 
}
\label{fig:visualization1}
\end{figure}

\begin{figure}[htbp]
\centering
\includegraphics[width=0.75\linewidth]{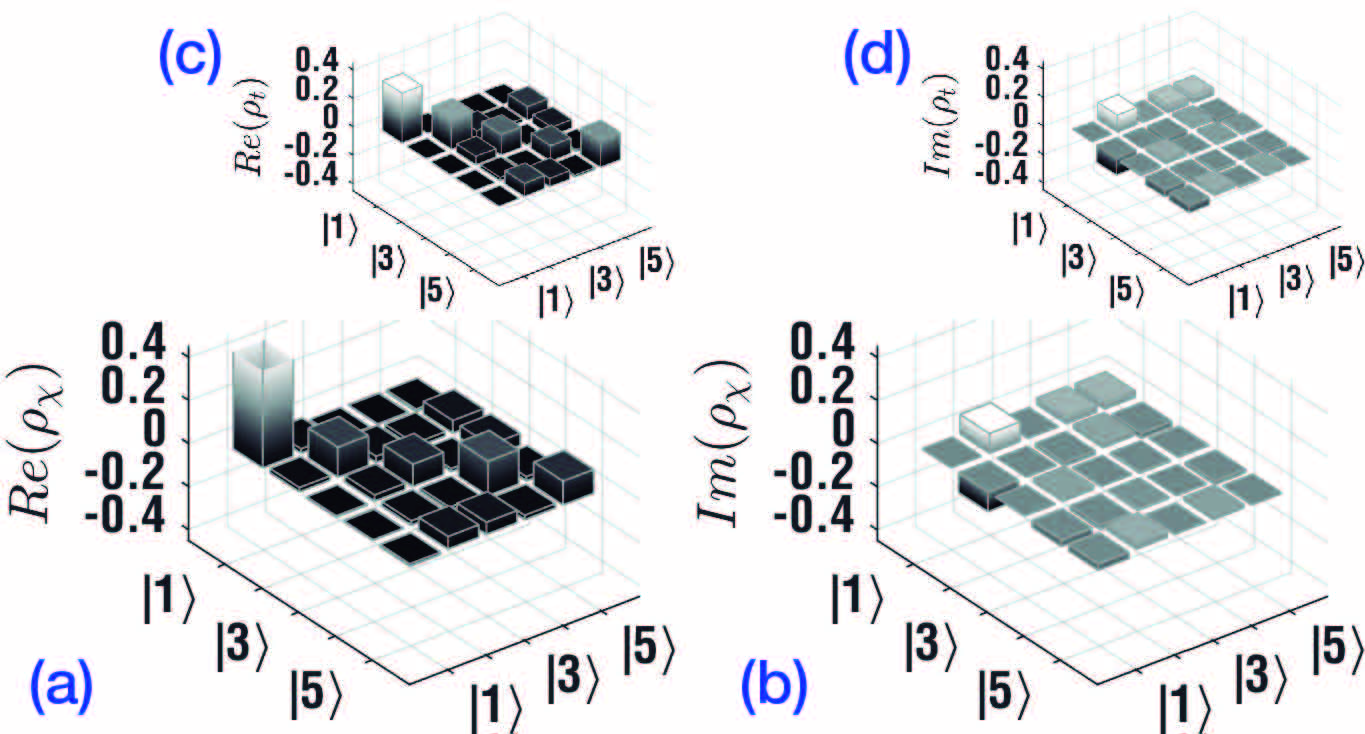}
\caption{{\it Amplitude-Phase-shift} (APS): Predicted density matrices $\rho_{out}=\mathcal{E}_{AS}(\rho_{in})$, for an uniform-respect-to-$|0\rangle$ APS, in $d=5$. Plots (c) and (d) show the real and imaginary parts of the density matrices of the outputs states, obtained by means of the Kraus decomposition of the process, 
$\mathcal{E}_{APS}$. Plots (a) and (b) are the real and imaginary parts corresponding to the density matrices obtained by means of the reconstructed matrix of such a process, $\chi$, after SQPT. The input state is a random state. 
}
\label{fig:visualization2}
\end{figure}

\bibliography{sample}


\end{document}